\documentclass{appolb}
\usepackage{epsfig}
\begin{document}

\title{TOROIDAL SUPER-HEAVY NUCLEI IN SKYRME-HARTREE-FOCK APPROACH\thanks{
Presented at the Zakopane Conference on Nuclear Physics,
September 1-7, 2008}}
\author{A. Staszczak
\address{Institute of Physics, Maria Curie-Sk{\l}odowska University, Lublin, Poland
\\
Department of Physics, University of Tennessee, Knoxville, USA
\\
Physics Division, Oak Ridge National Laboratory, Oak Ridge, USA}
\and C. Y. Wong
\address{Physics Division, Oak Ridge National Laboratory, Oak Ridge, USA}
}
\maketitle

\begin{abstract}
Within the self-consistent constraint Skyrme-Hartree-Fock+BCS model
(SHF+BCS), we found equilibrium toroidal nuclear density distributions
in the region of super-heavy elements. For nuclei with a sufficient
oblate deformation ($Q_{20}\leq$ -200 b), it becomes energetically
favourable to change the genus of nuclear surface from 0 to 1, i.e.,
to switch the shape from a biconcave disc to a torus. The energy of
the toroidal (genus=1) SHF+BCS solution relative to the compact
(genus=0) ground state energy is strongly dependent both on the atomic
number Z and the mass number A. We discuss the region of Z and A where
the toroidal SHF+BCS total energy begins to be a global minimum.
\end{abstract}


\section{Introduction}

The term \emph{doughnut nuclei} was coined by J. A. Wheeler
\cite{Whee50,Euwe} many years ago. In 1970's and 80's the idea of a
toroidal geometry in nuclear physics was investigated in a framework
of the liquid drop model (LDM) and shell corrections
\cite{Wong72,Wong73,Wong79,Wong78,Wong85}. In 1990's
\cite{Bauer92,Moret92,Moret93,Xu93,Xu94,Handz95} and more recently
\cite{Soch08,Soch08a} the toroidal breakup configurations were found
in the simulations of heavy-ion collisions using the
Boltzmann-Nordheim-Vlasov (BNV) or the Boltzmann-Uehling-Uhlenbeck
(BUU) kinetic transport models. The stability of toroidal nuclei
against a change of the quadrupole moment was studied recently in the
Hartree-Fock-Bogolibov (HFB) theory \cite{Ward07} with the Gogny D1S
force \cite{Berg85} and in the semiclassical extended Thomas-Fermi
(ETF) method \cite{Vinas08} with the Skyrme SkM* force \cite{Bart85}.

\begin{figure}
\centerline{\epsfig{file=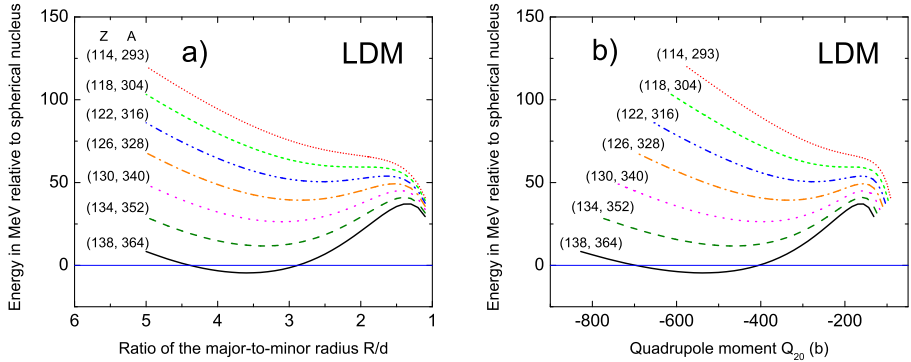,width=12cm,angle=0}}
%
\caption{a) Potential energy curves for toroidal (genus=1) nuclei of
different atomic numbers Z and mass numbers A as a function of the
aspect ratio $R/d$, where $R$ is the major radius and $d$ the minor
radius of the toroid. b) The same as a function of the quadrupole
moment $Q_{20}$ defined in Eq.\ (\ref{eqn:1}).}
\label{fig1}
\end{figure}

The purpose of this work is to study the stability of toroidal
super-heavy nuclei against collective deformations within the
self-consistent constraint Skyrme-Hartree-Fock+BCS framework
(SHF+BCS) with the Skyrme SkM* force \cite{Bart85}. Since the
classical LDM and semiclassical ETF approaches do not include shell
effects, we would like to examine quantitatively the role of the
shell effects in the description of toroidal nuclei.

The HF+BCS equations were solved using the code HFODD
\cite{Dob97,Dob04} that uses the basis expansion method in a
three-dimensional Cartesian deformed HO basis. The details of our
model are the same as in Ref.~\cite{Stas07}.

\section{Results}
In topology, the genus of a closed surface is equal to the number of
handles on the surface. For instance, a torus is a surface with
genus=1 while a sphere or disc has genus=0. The shape of a toroidal
nucleus can be characterized by an aspect ratio $R/d$, where the major
radius $R$ is measured from the center of the toroid to the center of
the circular meridian and the minor radius $d$ is the radius of the
meridian.  Assuming axial symmetry and a uniform mass distribution we
can calculate a quadrupole moment $Q_{20}$ for a given aspect ratio
$R/d$ as
\begin{eqnarray}
Q_{20} &=& \sqrt{\frac{16\pi}{5}} \int d^3 r
           \varrho(\vec{r}) r^2 Y_{20} (\cos\vartheta) \label{eqn:1} \\
       &=& \frac{A}{4}{R_0 \!\!\!\!\! {}^{{}^{{}^{\circ}}}}^2\left(\frac{2}{3\pi \cosh\eta_0}
           \right)^{2/3} \left[\sinh^2\eta_0 -
           5\cosh^2\eta_0\right], \nonumber
\end{eqnarray}
where in the toroidal coordinates \cite{Wong73}
$\varrho(r)=\frac{A}{V}\Theta(\eta-\eta_{0})$, $\cosh\eta_0=R/d$ and
$R_0 \!\!\!\!\! {}^{{}^{{}^{\circ}}}$ ~is a spherical radius of the same volume
$V=\left(4\pi/3\right){R_0 \!\!\!\!\! {}^{{}^{{}^{\circ}}}}^3 =
2\pi^2Rd^2$.

Figure~\ref{fig1} shows the potential energy curves for toroidal
nuclei with 114$\leq$Z$\leq$138 calculated within the LDM
\cite{Wong73,Wong79}. The potential energy curves are plotted as a function
of $R/d$ and $Q_{20}$ in the panel a) and b), respectively. As one
can see all toroidal super-heavy nuclei (genus=1) show an oblate
deformation with $Q_{20}\leq$ -200 b. In addition, as the atomic
number Z exceeds 138, the genus=1 equilibrium lies at an energy even
lower than that for the spherical shape.

\begin{figure}
\centerline{\epsfig{file=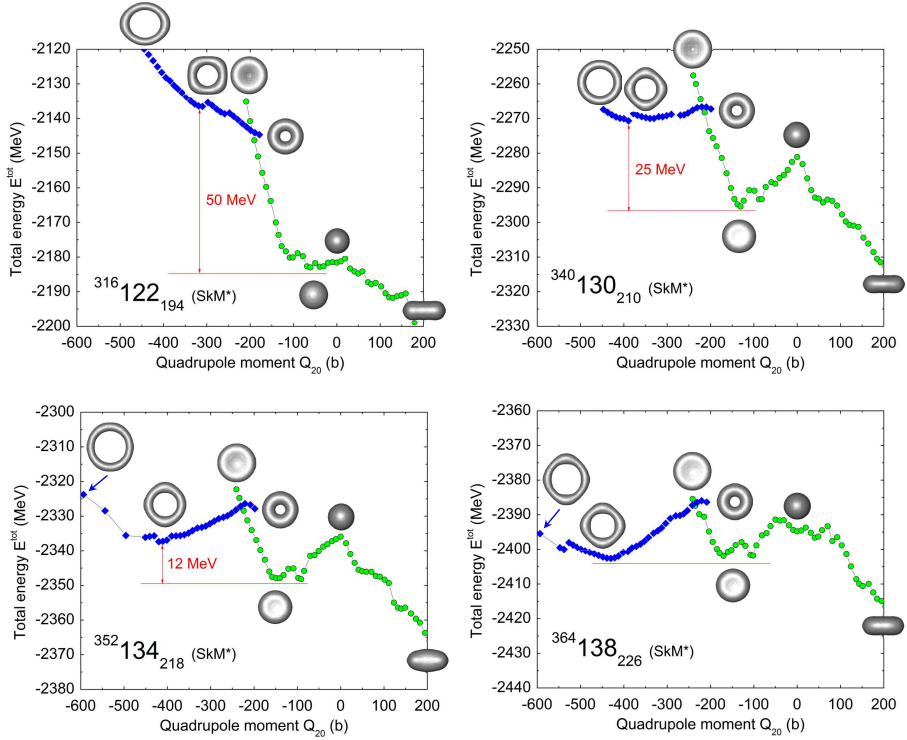,width=12cm,angle=0}}
%
\caption{The total binding energy ($E^{tot}$) of the toroidal
(genus=1) and compact (genus=0) SHF+BCS solutions as a function of
the quadrupole moment $Q_{20}$ for $^{316}122$, $^{340}130$,
$^{352}134$ and $^{364}138$. The self-consistent density
distributions with the compact and toroidal topologies are also shown,
where a z-axis lies on the surface of the page for a prolate
deformation and is perpendicular to the page in the case of an
oblate deformation.}
\label{fig2}
\end{figure}

In Fig.~\ref{fig2} we display the total binding energy ($E^{tot}$)
calculated within SHF+BSC approach as a function of the quadrupole
moment $Q_{20}$ for $^{316}122$, $^{340}130$, $^{352}134$ and
$^{364}138$ super-heavy nuclei. The two branches of $E^{tot}$
corresponding to a compact (genus=0) and a toroidal (genus=1)
SHF+BCS solutions have been found. Both solutions coexist in the
vicinity of $Q_{20}\simeq$ -200 b. But with the further increase of
the oblate deformation it becomes energetically favourable to change
the genus of nuclear surface from 0 to 1, i.e., to switch the shape
from a biconcave disc to a torus. For $Q_{20}<$ -250 b only genus=1
self-consistent solutions exist.

We also show the self-consistent density distributions obtained for
both branches with the compact and toroidal topology. Since the
SHF+BCS model is not restricted to the axial symmetry, the
toroidal density distributions show the characteristic sausage
deformations \cite{Wong73}. The sausage instability is responsible
for a multifragmentation of toroidal nuclei and in the case of
$^{316}122$ and $^{340}130$ the toroidal solutions cease to exist at
$Q_{20}\simeq$ -450 b.

Similarly to the LDM results showed in Fig.~\ref{fig1}, the genus=1
and genus=0 total energy minima become closer in energy with
increasing atomic number Z and mass number A. For Z=138 and
A=364 the toroidal equilibrium again begins to be the global
minimum.

The neutron Fermi energies ($\lambda^n$) oscillate around -6 MeV with
a change of quadrupole deformation for all nuclei from
Fig.~\ref{fig2}. The proton Fermi energies ($\lambda^p$) stronger
depend on $Q_{20}$. For the prolate deformations $\lambda^p$ oscillate
between -2 and -1 MeV but in the region of oblate deformation
$\lambda^p$ decreases to the value of $\lambda^n$ at $Q_{20}\simeq$
-400 b. Thus, the toroidal SHF+BCS solutions are more stable against
$\beta$-decay than the compact (genus=0) solutions.


In conclusion, it appears that our self-consistent SHF+BCS
calculations for the toroidal nuclei are consistent with the
classical LDM \cite{Wong79} and semiclassical ETF \cite{Vinas08}
models as well as with the HFB theory \cite{Ward07}.

The toroidal nuclear density distribution is not a surprising
phenomenon, but a regular characteristic of the strongly oblate
deformed heavy and super-heavy nuclei. The doughnut shaped nuclei are
not the exception but rather the norm in this region of deformations!

One of us (A.S) would like to thank M. Warda for interesting
discussions which led to this investigation.
This work was supported in part by the National Nuclear
Security Administration under the Stewardship Science Academic
Alliances program through the U.S. Department of Energy Research
Grant DE-FG03-03NA00083; by the U.S. Department of Energy under
Contract Nos.\ DE-FG02-96ER40963 (University of Tennessee),
DE-AC05-00OR22725 with UT-Battelle, LLC (Oak Ridge National
Laboratory), and DE-FC02-07ER41457 (UNEDF SciDAC Collaboration); by
the Polish Ministry of Science and Higher Education under Contract
No.~N202~179~31/3920.

\end{document}